# The Hubble Legacy Fields (HLF-GOODS-S) v1.5 Data Products: Combining 2442 Orbits of GOODS-S/CDF-S Region ACS and WFC3/IR Images.

*Garth Illingworth[1], Daniel Magee[1], Rychard Bouwens[2], Pascal Oesch[3,] Ivo Labbe[2], Pieter van Dokkum[3], Katherine Whitaker[4], Bradford Holden[1], Marijn Franx[2], and Valentino Gonzalez[5]*

## Abstract

We have submitted to MAST the 1.5 version data release of the Hubble Legacy Fields (HLF) project covering a 25 x 25 arcmin area over the GOODS-S (ECDF-S) region from the HST archival program AR-13252. The release combines exposures from Hubble's two main cameras, the Advanced Camera for Surveys (ACS/WFC) and the Wide Field Camera 3 (WFC3/IR), taken over more than a decade between mid-2002 to the end of 2016. The HLF includes essentially all optical (ACS/WFC F435W, F606W, F775W, F814W and F850LP filters) and infrared (WFC3/IR F098M, F105W, F125W, F140W and F160W filters) data taken by Hubble over the original CDF-S region including the GOODS-S, ERS, CANDELS and many other programs (31 in total). The data has been released at https://archive.stsci.edu/prepds/hlf/ as images with a common astrometric reference frame, with corresponding inverse variance weight maps. We provide one image per filter of WFC3/IR images at 60 mas per pixel resolution and two ACS/WFC images per filter, at both 30 and 60 mas per pixel. Since this comprehensive dataset combines data from 31 programs on the GOODS-S/CDF-S, the AR proposal identified the MAST products by the global name "Hubble Legacy Field", with this region being identified by "HLF-GOODS-S". This dataset complements that of the Frontier Fields program. The total incorporated in the HLF-GOODS-S is 5.8 Msec in 7211 exposures from 2442 orbits. This is ~70% of a HST full cycle!

## Introduction

The GOODS-S/CDF-S region has **accumulated** a uniquely large amount of data from the Advanced Camera for Surveys (ACS/WFC) and the Wide Field Camera 3 (WFC3/IR) over more than a decade of observations from 2002 to 2016. While the major datasets like GOODS-S (Giavalisco et al., 2004), ERS and CANDELS (Koekemoer et al., 2011) have been made available individually as high level science data products in MAST, the data from numerous other programs has not been readily available, nor has that data been combined with the major datasets. A total of 30 HST programs have data from the ACS and WFC3/IR that on this region. A first major combination of disparate datasets in this region was made on the HUDF when all data on that area through early 2013, both ACS and WFC3/IR, following the completion of the HUDF12 program, were combined into the HUDF/XDF and submitted to MAST as a high level science data products (Illingworth et al., 2013). The lack of a combined dataset on the whole CDF-S region has been unfortunate given the HST resources that have been expended on this area, and its central role in so many studies of distant galaxies.

---

[1] UCO/Lick Observatory, University of California, Santa Cruz,

[2] Leiden Observatory, Leiden University

[3] Department of Astronomy, Yale University

[4] Department of Astronomy, University of Massachusetts Amherst

[5] Department of Astronomy, Universidad de Chile

Given this, an Archival program AR-13252 was submitted and approved to carry out the full combination of all the datasets on GOODS-S/CDF-S. Since this dataset combines all images in the HST archive on the GOODS-S/CDF-S to date, the archival proposal identified the data product under the global name "Hubble Legacy Field," with the designation HLF-GOODS-S for this field. The HLF includes all the data from the GOODS-South, CANDELS-S, ERS, ECDF-S, HUDF, HUDF parallels, and numerous other programs (tabulated below in Table 2).

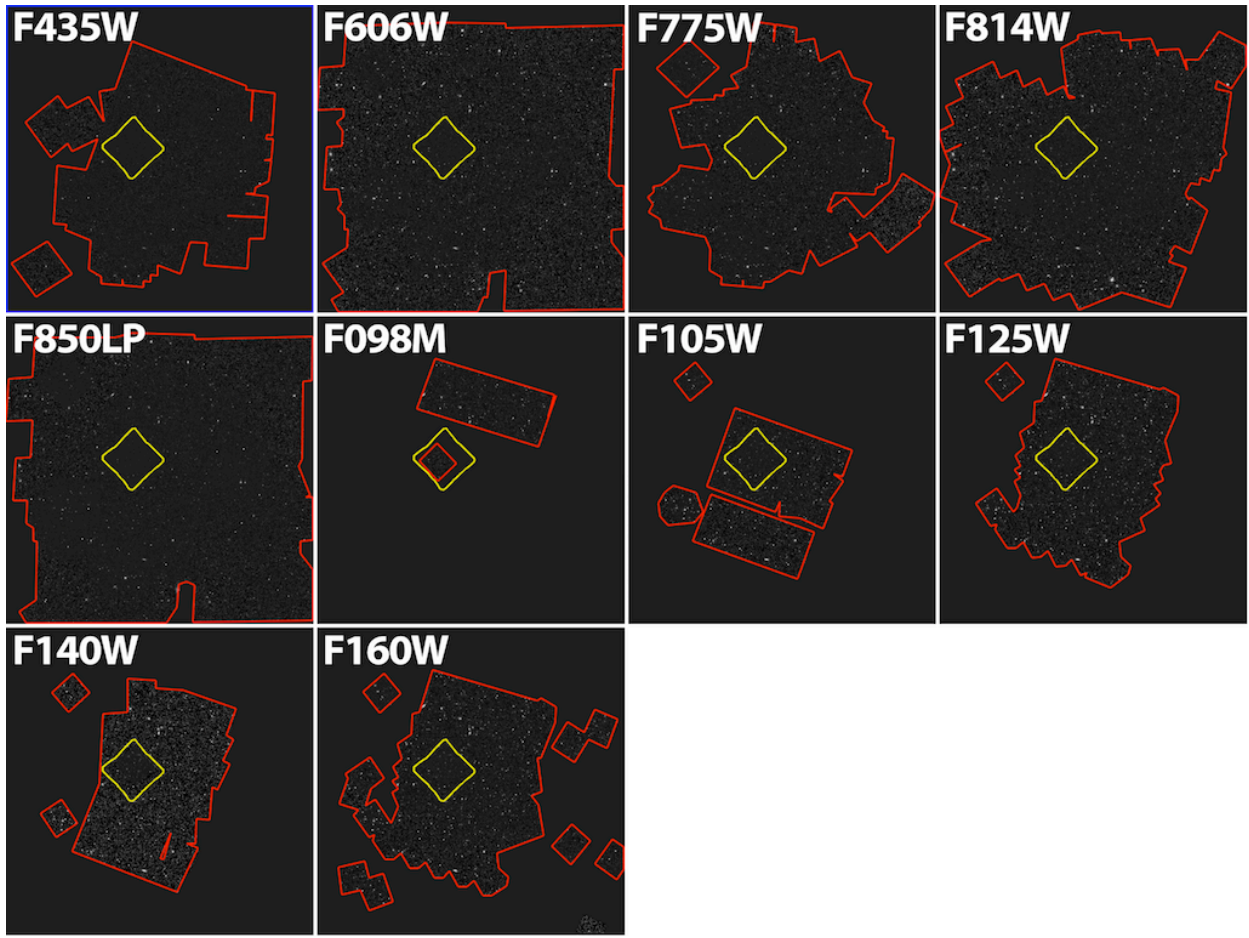

**Figure 1.** The HLF-GOODS-S dataset footprints for the five ACS/WFC and five WFC/IR filters are shown in red. The footprint of the HUDF/XDF dataset is show in yellow.

Given the large area, a global astrometric solution had to be bootstrapped from the smaller datasets. All the image mosaics have been produced using the same tangent point as the original GOODS-S dataset. The HLF includes the ACS/WFC optical filters (F435W, F606W, F775W, F814W and F850LP) and WFC3/IR infrared filters (F098M, F105W, F125W, F140W and F160W). See also http://firstgalaxies.org/hlf for future updates and further information.

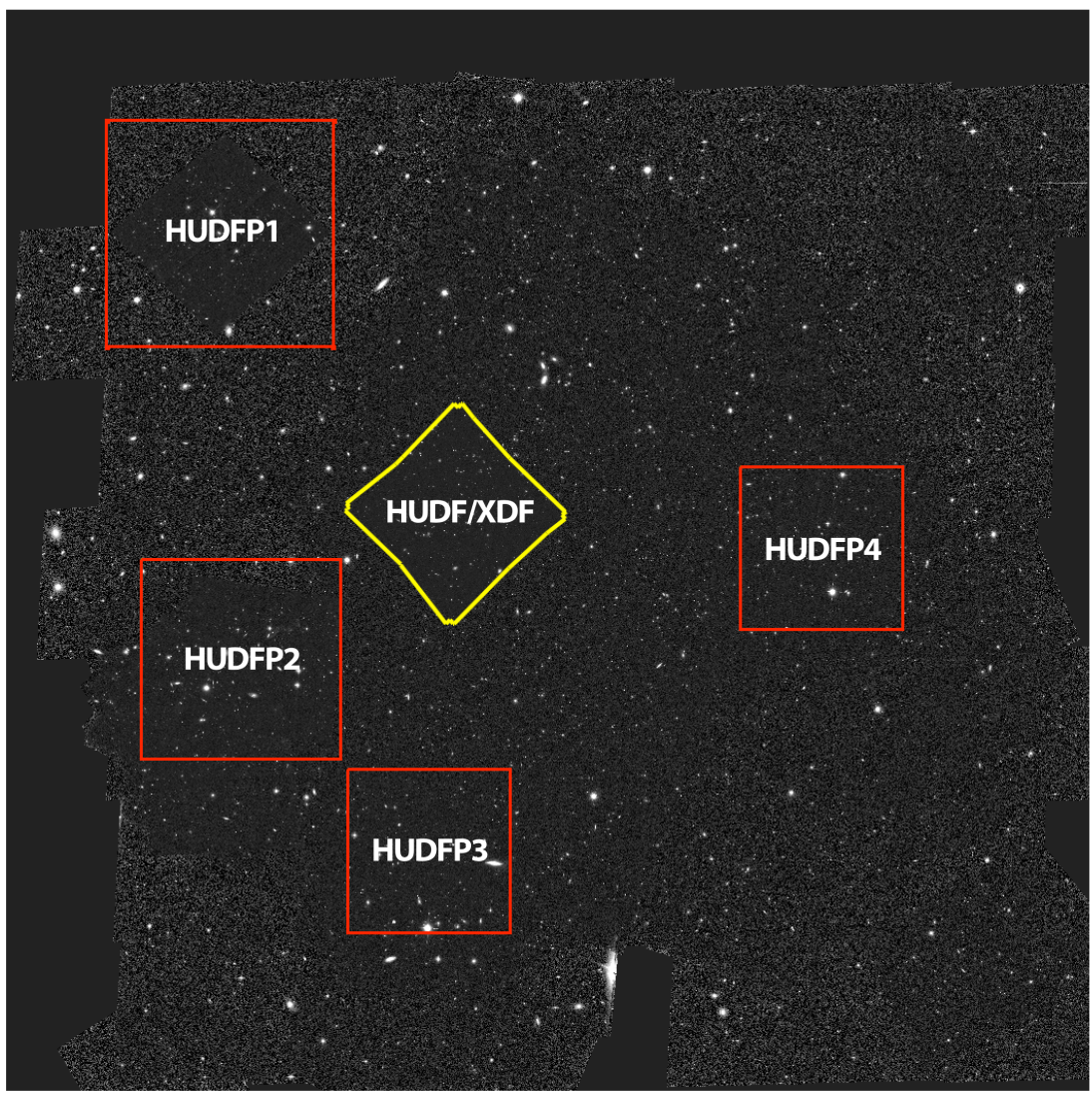

**Figure 2.** The five deep areas in the HLF-GOODS-S. The deep area cutouts HLF-HUDFP1, HLF-HUDFP2, HLF-HUDFP3 and HLF-HUDFP4 footprints are shown in red. The footprint of the HUDF/XDF dataset is show in yellow.

The HLF program complements the Frontier Fields program (Lotz et al. 2016) by making available a complete aligned optical and IR dataset taken on Hubble's most-imaged field region.

### HLF Dataset for the GOODS-South region (HLF-GOODS-S)

The datasets used in the HLF-GOODS-S delivery to MAST are shown in Figures 1-3. The 1.5 version of the HLF dataset provides ACS/WFC and WFC3/IR images covering a 25 x 25 arcmin area (Figure 1); as well as smaller cutouts of four deep areas in the GOODS-South region (Figure 2). All the data covering the original HUDF field within the GOODS-South area was previously released as high level science product - the HUDF/XDF field (Illingworth et al. 2013). The HLF-GOODS-S dataset currently includes five ACS/WFC filters (F435W, F606W, F775W, F814W, F850LP) and five WFC3/IR filters (F098M, F105W, F125W, F140W & F160W). An exposure time map is shown in Figure 3 illustrating the depth in each of the ten filters.

### Citation

Please reference "Illingworth, Magee, Bouwens, Oesch et al, 2017, in preparation".

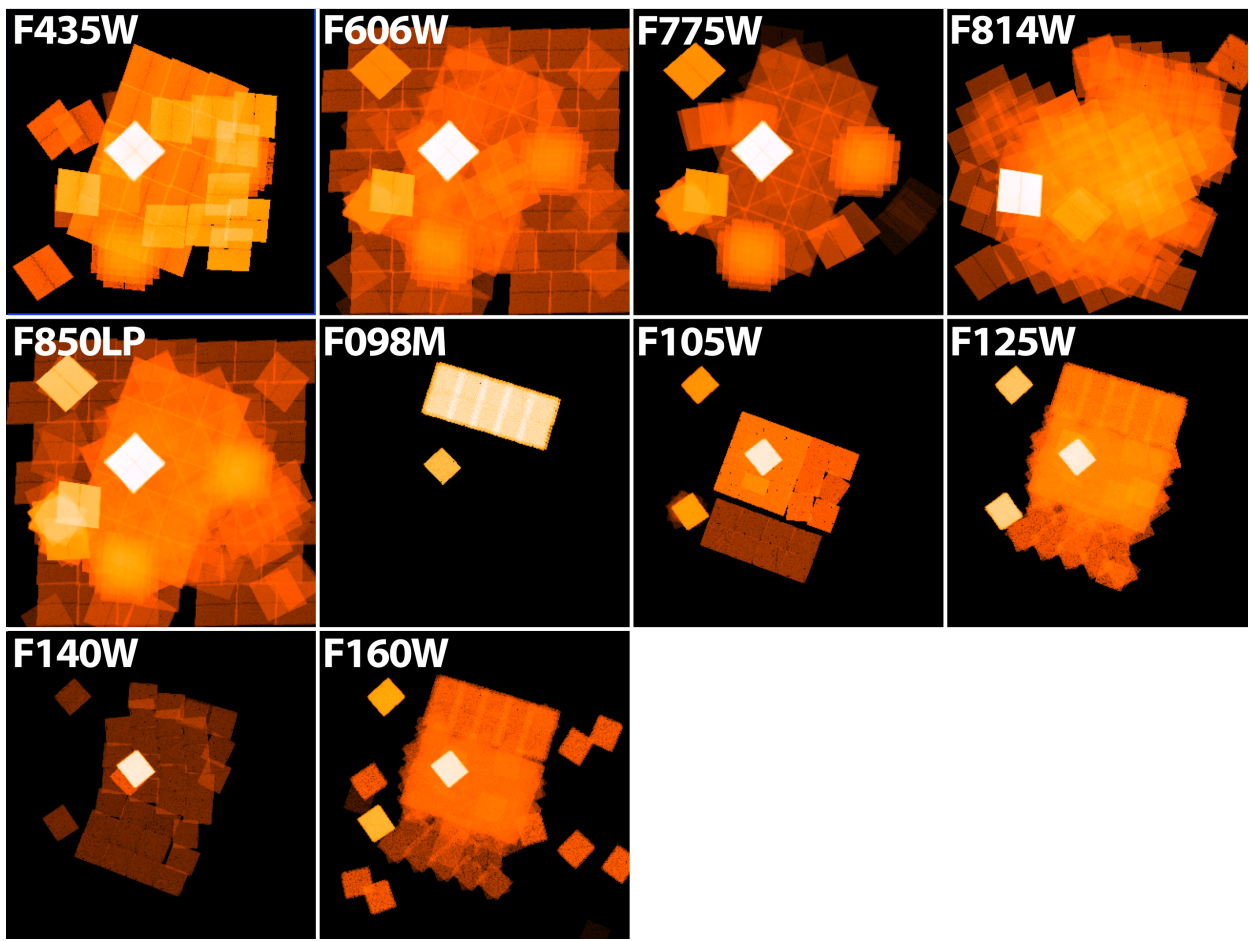

**Data Products and Links**

The HLF v1.5 data products are accessible from the Milkulski Archive for Space Telescope (MAST) High Level Science Products pages at:

https://archive.stsci.edu/prepds/hlf/

The released dataset includes fully reduced, science ready images (*_sci.fits) together with the associated weight maps (*_wht.fits) at 30mas/pixel (for ACS only) and 60mas/pixel (for ACS+WFC3 images).

The data are organized into sets of images by passband (ACS/WFC: F435W, F606W, F775W, F814W & F850LP, and WFC3/IR: F098M, F105W, F125W, F140W & F160W) and image scale. Each 60 mas/pixel HLF-GOODS-S image is 25k x 25k pixels and each 30 mas/pixel image is 50k x 50k pixels. For each filter we provide the drizzled science image and a weight image. All data use the same tangent point as the original GOODS-S dataset (*R.A. = 53.122751, Dec. = –27.805089 J2000*). Previous v0.5 and v1.0 data releases were made available in July 2015 and April 2016, respectively. The v1.5 supersedes these older dataset due to improvements in the data processing and additional data.

All images used a drizzle *pixfrac* parameter value of 0.8 (*final_pixfrac=0.8*). The weight map image is equal to the inverse variance (i.e., *$1/rms^2$*) per pixel. A detailed discussion of weight map conventions and noise correlation in drizzling, can be found in Casertano et al., 2000, especially in their Section 3.5 and Appendix A.

All exposures used to produce the HLF were first visually inspected to identify any data quality issues (loss of guiding, excessive background, pointing accuracy) and any image which could not be corrected was rejected for processing. During this visual inspection we also identify images affected by satellite trails and optical ghosts from filter reflections generated by bright stars (ACS). We updated the data quality array to ensure these artifacts were masked during final processing. ACS/WFC data taken after HST Servicing Mission 4 (SM4) were correct for charge transfer efficiency (CTE) degradation.

A summary of the exposure times and number of exposures for each filter is shown in Table 1. The total incorporated in the HLF-GOODS-S is 5.8 Msec in 7211 exposures from 2442 orbits. This is ~70% of a HST full cycle!

## Table 1. Exposure Summary

Guys:

| Filter | Exposure Time (s) | Number of Exposures |
|---|---:|---:|
| F435W | 448,488 | 443 |
| F606W | 537,944 | 712 |
| F775W | 770,190 | 849 |
| F814W | 840,211 | 1,288 |
| F850LP | 1,495,092 | 1,902 |
| F098M | 53,499 | 68 |
| F105W | 484,734 | 429 |
| F125W | 440,755 | 598 |
| F140W | 115,590 | 218 |
| F160W | 596,277 | 704 |
| **Totals** | **5,861,244** | **7,211** |

## Observations

Observations included in the v1.5 version of the HLF-GOODS-S region were taken from July 2002 to October 2016 from 31 different HST programs (Table 2).

**Table 2. Programs used for the HFL-GOODS-S region**

| Program ID | Program Title | Program PI |
|---|---|---|
| 9352 | The Deceleration Test from Treasury Type Ia Supernovae at Redshifts 1.2 to 1.6 | Adam Riess |
| 9425 | The Great Observatories Origins Deep Survey: Imaging with ACS | Mauro Giavalisco |
| 9480 | Cosmic Shear With ACS Pure Parallels | Jason D. Rhodes |
| 9488 | Cosmic Shear - with ACS Pure Parallel Observations | Kavan Ratnatunga |
| 9500 | The Evolution of Galaxy Structure from 10,000 Galaxies with 0.1<z<1.2 | Hans-Walter Rix |
| 9575 | ACS Default {Archival} Pure Parallel Program | William B. Sparks |
| 9793 | The Grism-ACS Program for Extragalactic Science {GRAPES} | Sangeeta Malhotra |
| 9803 | Deep NICMOS Images of the UDF | Rodger I. Thompson |
| 9978 | The Ultra Deep Field with ACS | Steven Beckwith |
| 9984 | Cosmic Shear With ACS Pure Parallels | Jason D. Rhodes |
| 10086 | The Ultra Deep Field with ACS | Steven Beckwith |
| 10189 | PANS-Probing Acceleration Now with Supernovae | Adam Riess |
| 10258 | Tracing the Emergence of the Hubble Sequence Among the Most Luminous and Massive Galaxies | Claudia Kretchmer |

| Program ID | Program Title | Program PI |
|---|---|---|
| 10340 | PANS | Adam Riess |
| 10530 | Probing Evolution And Reionization Spectroscopically {PEARS} | Sangeeta Malhotra |
| 10632 | Searching for galaxies at z>6.5 in the Hubble Ultra Deep Field | Massimo Stiavelli |
| 11144 | Building on the Significant NICMOS Investment in GOODS: A Bright, Wide-Area Search for z>=7 Galaxies | Rychard Bouwens |
| 11359 | Panchromatic WFC3 survey of galaxies at intermediate z: Early Release Science program for Wide Field Camera 3. | Robert W. O'Connell |
| 11563 | Galaxies at z~7-10 in the Reionization Epoch: Luminosity Functions to <0.2L* from Deep IR Imaging of the HUDF and HUDF05 Fields | Garth D. Illingworth |
| 12007 | Supernova Followup | Garth D. Illingworth |
| 12060 | Cosmic Assembly Near-IR Deep Extragalactic Legacy Survey — GOODS-South Field, Non-SNe-Searched Visits | Sandra M. Faber |
| 12061 | Cosmic Assembly Near-IR Deep Extragalactic Legacy Survey — GOODS-South Field, Early Visits of SNe Search | Sandra M. Faber |
| 12062 | Galaxy Assembly and the Evolution of Structure over the First Third of Cosmic Time - III | Sandra M. Faber |
| 12099 | Supernova Follow-up for MCT | Adam Riess |
| 12177 | 3D-HST: A Spectroscopic Galaxy Evolution Treasury | Pieter Van Dokkum |
| 12461 | Supernova Follow-up for MCT | Adam Riess |
| 12498 | Did Galaxies Reionize the Universe? | Richard S. Ellis |
| 12866 | A Morphological Study of ALMA Identified Sub-mm Galaxies with HST/WFC3 | Mark Swinbank |

| Program ID | Program Title | Program PI |
|---|---|---|
| 12990 | Size Growth at the Top: WFC3 Imaging of Ultra-Massive Galaxies at 1.5 < z < 3 | Adam Muzzin |
| 13779 | The Faint Infrared Grism Survey (FIGS) | Sangeeta Malhotra |
| 13872 | The GOODS UV Legacy Fields: A Full Census of Faint Star-Forming Galaxies at z~0.5-2 | Pascal Oesch |

## Acknowledgements

The Hubble Legacy Fields program, supported through AR-13252, is based on observations made with the NASA/ESA Hubble Space Telescope, obtained at the Space Telescope Science Institute, which is operated by the Association of Universities for Research in Astronomy, Inc., under NASA contract NAS 5-26555. Financial support for this program is gratefully acknowledged. The Hubble datasets used in this program are from numerous programs that are all listed in Table 2 above, and were taken for our analysis from the MAST archive. We thank all of those who programs were combined into the HLF-GOODS-S for providing a set of extraordinary data that will have legacy value for the community into the JWST era and beyond.

## References

Casertano, S., de Mello, D., Dickinson, M., et al., 2000, AJ, 120, 2747.

Giavalisco et al., 2004, ApJ, 600, L93.

Illingworth, G. D., Magee, D., Oesch, P. A., Bouwens, R., et al. 2013, ApJS, 209.

Koekemoer et al., 2011, ApJS, 197, 36.

Lotz et al., 2016, arXiv:1605.06567.